\documentclass[12pt,a4paper]{iopart}
\usepackage{graphicx}
\usepackage{multirow}
\usepackage{acronym}
\usepackage[dvipsnames]{xcolor}

\newcommand{\eqref}[1]{\eref{#1}}

\newcommand{\coloredtext}[1]{{#1}}
\newcommand{\hgen}{$\eta$Gen}
\newcommand{\etagen}{EtaGen}
\newcommand{\utrg}{\coloredtext{excess}}
\newcommand{\autrg}{\coloredtext{an excess}}

\newcommand{\utrgs}{\coloredtext{excesses}}
\newcommand{\trg}{\coloredtext{event trigger}}
\newcommand{\atrg}{\coloredtext{an event trigger}}
\newcommand{\Atrg}{\coloredtext{An event trigger}}
\newcommand{\trgs}{\coloredtext{event triggers}}

\newcommand{\firstgw}{GW150914}
\newcommand{\secondgw}{GW151226}

\newcommand{\thirdgw}{GW170104}
\newcommand{\fourthgw}{GW170608}
\newcommand{\fifthgw}{GW170814}
\newcommand{\sixthgw}{GW170817}

\newcommand{\numdata}[1]{{#1}}
\newcommand{\snrth}{\numdata{5.5}}
\newcommand{\usnrth}{\numdata{2.75}}
\newcommand{\madratio}{\numdata{1.48}}

\newcommand{\allpercent}{\numdata{95.47\,\%}}
\newcommand{\alltacc}{\numdata{$8.32 \times 10^{-3}$\,s}}
\newcommand{\allfacc}{\numdata{381.07\,Hz}}

\newcommand{\sgtotal}{\numdata{3132}}

\newcommand{\sgpercent}{\numdata{99.74\,\%}}
\newcommand{\sgtacc}{\numdata{$4.49 \times 10^{-3}$\,s}}
\newcommand{\sgfacc}{\numdata{223\,Hz}}

\newcommand{\wnbtotal}{\numdata{3121}}

\newcommand{\wnbpercent}{\numdata{91.19\,\%}}
\newcommand{\wnbtacc}{\numdata{$1.11 \times 10^{-2}$\,s}}
\newcommand{\wnbfacc}{\numdata{497\,Hz}}

\begin{document}

\title[Generating Event Triggers Based on HHT and Its Application to GW Data]{Generating Event Triggers Based on Hilbert-Huang Transform and Its Application to Gravitational-Wave Data}

\author{Edwin J. Son$^{1}$, Whansun Kim$^{1}$, Young-Min Kim$^{2}$, Jessica McIver$^{3}$, John J. Oh$^{1}$ and Sang Hoon Oh$^{1}$}
\address{$^1$ Division of Basic Researches for Industrial Mathematics,
National Institute for Mathematical Sciences, Daejeon 34047, Republic of Korea}
\address{$^2$ School of Natural Science, Ulsan National Institute of Science and Technology, Ulsan 44919, Republic of Korea}
\address{$^3$ LIGO, California Institute of Technology, Pasadena, CA
91125, USA}
\eads{\mailto{eddy@nims.re.kr}, \mailto{iou78@nims.re.kr}, \mailto{ymkim715@unist.ac.kr}, \mailto{jlmciver@caltech.edu}, \mailto{johnoh@nims.re.kr}, \mailto{shoh@nims.re.kr}}

\vspace{10pt}
\begin{indented}
\item[]\today
\end{indented}

\begin{abstract}
We present a new event trigger generator based on the Hilbert-Huang transform, named EtaGen ($\eta$Gen).
It decomposes a time-series data into several adaptive modes without imposing \textit{a priori} bases on the data.
The adaptive modes are used to find transients (\utrgs{}) in the background noises.
A clustering algorithm is used to gather \utrgs{} corresponding to a single event and to reconstruct its waveform.
The performance of EtaGen is evaluated by how many injections in the LIGO simulated data are found.
EtaGen is viable as an event trigger generator when compared directly with the performance of Omicron, which is currently the best event trigger generator used in the LIGO Scientific Collaboration and Virgo Collaboration.
\end{abstract}

%
\vspace{2pc}

%
%
%

\acrodef{gw}[GW]{gravitational-wave}
\acrodef{ligo}[LIGO]{Laser Interferometer Gravitational-wave Observatory}
\acrodef{aligo}[aLIGO]{Advanced \ac{ligo}}
\acrodef{adv}[AdV]{Advanced Virgo}
\acrodef{lsc}[LSC]{\ac{ligo} Scientific Collaboration}
\acrodef{lvc}[LVC]{\ac{lsc}-Virgo Collaboration}
\acrodef{imf}[IMF]{intrinsic mode function}
\acrodef{emd}[EMD]{empirical mode decomposition}
\acrodef{semd}[SEMD]{sliding \ac{emd}}
\acrodef{hht}[HHT]{Hilbert-Huang transform}
\acrodef{snr}[SNR]{signal-to-noise ratio}
\acrodef{std}[STD]{standard deviation}
\acrodef{mad}[MAD]{median absolute deviation}

\acresetall
\section{Introduction}
Recently, the \ac{lsc} and Virgo collaboration announced the observations of confirmed \ac{gw} signals (\firstgw{}, \secondgw{}, \thirdgw{}, \fourthgw{}, \fifthgw{}, and \sixthgw{}) from the binary black hole mergers and a binary neutron star merger~\cite{TheLIGOScientific:2016wyq, Abbott:2016nmj, Abbott:2017vtc, Abbott:2017gyy, Abbott:2017oio, TheLIGOScientific:2017qsa}.
In order to make these discoveries possible, the \ac{lvc} characterizes the data to mitigate a high rate of transient noise that can mimic or bias the source property estimation of true \ac{gw} events~\cite{0264-9381-32-24-245005,TheLIGOScientific:2016zmo}. The event trigger generators are a critical tool for identifying and characterizing transient noise. Event trigger generators extract instances of excess power from a time series, called \emph{event triggers}.
Most of the event trigger generators in the \ac{lvc} impose wavelets as a basis set, which offer coarse time resolution at low frequencies~\cite{Huang:2005:book}. 


Here, we introduce an adaptive time-series analysis method, the \ac{hht}~\cite{Huang1998} to improve time resolution at low frequency, where there might be harmful noise sources such as earthquakes and light scattering as common problems that limit interferometer uptime and hinder the astrophysical searches. Moreover, in the third generation interferometer era like \emph{Cosmic Explorer}, the low-frequency band noise will become crucial in the major detector sensitivity.

The \ac{hht} consists of the \ac{emd} and the Hilbert spectral analysis. The \ac{emd} decomposes a time-series data into several adaptive modes\footnote{Here we describe a mode to be adaptive if it is determined during the decomposition process that depends on the original time series data as oppose to the mode that are determined by the predifined basis set.}, called \acp{imf}, without imposing a basis set.
Then, the Hilbert spectral analysis calculates the instantaneous amplitude and instantaneous frequency of each \acp{imf}.
It is a remarkable property of this adaptive approach that relieves the uncertainty principle in the signal processing~\cite{Huang:2005:book}.
The time resolution of each \ac{imf} is the same with that of the input data even in the low frequency band.
In this respect and by virtue of its adaptive nature, the \ac{hht} has been extensively used in the various area such as biomedical application of Electroencephalography data, financial analysis, geophysical and meteorological data analysis, etc (for more references, see~\cite{Huang2008}).
In particular, it has been introduced and studied for \ac{gw} signal search and noise detection in~\cite{Camp:2007ee,Stroeer:2009zz,Stroeer:2009hv,Kaneyama:2016fww,Sakai:2017ckm,Valdes:2017xce}.

In this paper, we propose a new event trigger generator based on the \ac{hht}, named \emph{\etagen{} (\hgen{})}.
The instantaneous amplitude of each \ac{imf} is used to mark the regions which excess a certain amplitude cut, and those regions are referred to as \emph{\utrgs{}}.
The \utrgs{} in all the \acp{imf} are characterized by their time and frequency informations, using the instantaneous frequency of the \ac{imf}.
The event triggers are then obtained by clustering nearby \utrgs{} in time-frequency plane.
Generated event triggers are used to diagnose correlations between transients in the \ac{gw} data and the detectors' behavior and environment, as well as to characterize transient noise that impacts the astrophysical searches.
One of byproduct of \etagen{} is to reconstruct the original waveform by simply adding the time-series corresponding to the \utrgs{} clustered in each event trigger.

This paper is organized as follows.
We first briefly review the \ac{hht} and introduce an online \ac{emd} method in section~\ref{sec:hht}.
Then, we describe the signal extraction and clustering algorithms of the \etagen{} followed by waveform reconstruction method in section~\ref{sec:etg}.
Then the performance of \etagen{} is found in section~\ref{sec:perf}.
Finally, conclusion and discussion are given in section~\ref{sec:dis}.

\begin{figure}[tbp]
\begin{center}
\setlength{\unitlength}{3em}
\begin{picture}(10.9,6.5)
  \put(2,6){\oval(2,1)}
  \put(1.5,5.9){START}
  \put(2,5.5){\vector(0,-1){.3}}
  \put(2,4.7){\oval(4,1)}
  \put(1.3,4.8){$s(t) := x(t),$}
  \put(1.6,4.4){$k := 1$}
  \put(2,4.2){\vector(0,-1){.3}}
  \put(2,2.9){\oval(4,2)}
  \put(.6,3.3){$\textrm{Find } u(t) \textrm{ and } \ell(t),$}
  \put(.6,2.8){$\textrm{the upper and lower}$}
  \put(.6,2.3){$\textrm{envelopes of } s(t)$}
  \put(2,1.9){\vector(0,-1){.3}}
  \put(2,.8){\oval(4,1.6)}
  \put(.6,1){$m(t) := \frac{u(t) + \ell(t)}{2},$}
  \put(.7,.4){$h(t) := s(t) - m(t)$}
  \put(4,.8){\vector(1,0){2.5}}
  \multiput(8.5,1.3)(2,-.5){2}{\line(-4,-1){2}}
  \multiput(8.5,1.3)(-2,-.5){2}{\line(4,-1){2}}
  \put(7.7,.7){$h(t) = \textrm{\acs{imf}}$ ?}
  \put(8.5,1.3){\line(0,1){.3}}
  \put(8.6,1.4){No}
  \put(8.5,1.6){\vector(-1,0){2}}
  \put(5.5,1.6){\oval(2,1)}
  \put(4.7,1.5){$s(t) := h(t)$}
  \put(5.5,2.1){\line(0,1){.5}}
  \put(5.5,2.6){\vector(-1,0){1.5}}
  \put(10.5,.8){\line(1,0){.3}}
  \put(10.15,1.1){Yes}
  \put(10.8,.8){\line(0,1){2}}
  \put(10.8,2.8){\vector(-1,0){.3}}
  \put(8.5,2.8){\oval(4,1.8)}
  \put(6.9,3){$c_k(t) := h(t),$}
  \put(6.9,2.4){$r(t) := x(t) - \sum_{j=1}^k c_j(t)$}
  \put(8.5,3.7){\vector(0,1){.3}}
  \multiput(8.5,5)(2,-.5){2}{\line(-4,-1){2}}
  \multiput(8.5,5)(-2,-.5){2}{\line(4,-1){2}}
  \put(8,4.4){STOP ?}
  \put(6.5,4.5){\line(-1,0){1.2}}
  \put(5.8,4.6){No}
  \put(5.3,4.5){\vector(0,-1){.2}}
  \put(5.3,3.7){\oval(2.4,1.2)}
  \put(4.5,3.8){$s(t) := r(t),$}
  \put(4.5,3.4){$k := k + 1$}
  \put(5.3,3.1){\line(0,-1){.2}}
  \put(5.3,2.9){\vector(-1,0){1.3}}
  \put(8.5,5){\vector(0,1){.5}}
  \put(8.7,5.1){Yes}
  \put(8.5,6){\oval(2,1)}
  \put(8.1,5.9){END}
\end{picture}
\caption{The flowchart of \acf{emd}. \acs{emd} does not use wavelets as a basis set. A segment of time-series data undergoes a process called \emph{sifting}: the segment is subtracted by the mean of its envelops to get an \acs{imf} (Step~\ref{emd:sifting} and \ref{emd:subtraction} of \acs{emd}). The remainder becomes an \acs{imf} if it satisfies a \acs{imf} criterion. Or the remainder undergoes sifting repeatedly until satisfying the criterion (Step~\ref{emd:imf}). Once an \acs{imf} is obtained, it is subtracted from the original segment (Step~\ref{emd:nextimf}) to obtain a residual. And the whole process is repeated on the residual until it meets the ending criterion (Step~\ref{emd}). Refer to the text for more detail.}
\label{fig:emd_flow}
\end{center}
\end{figure}
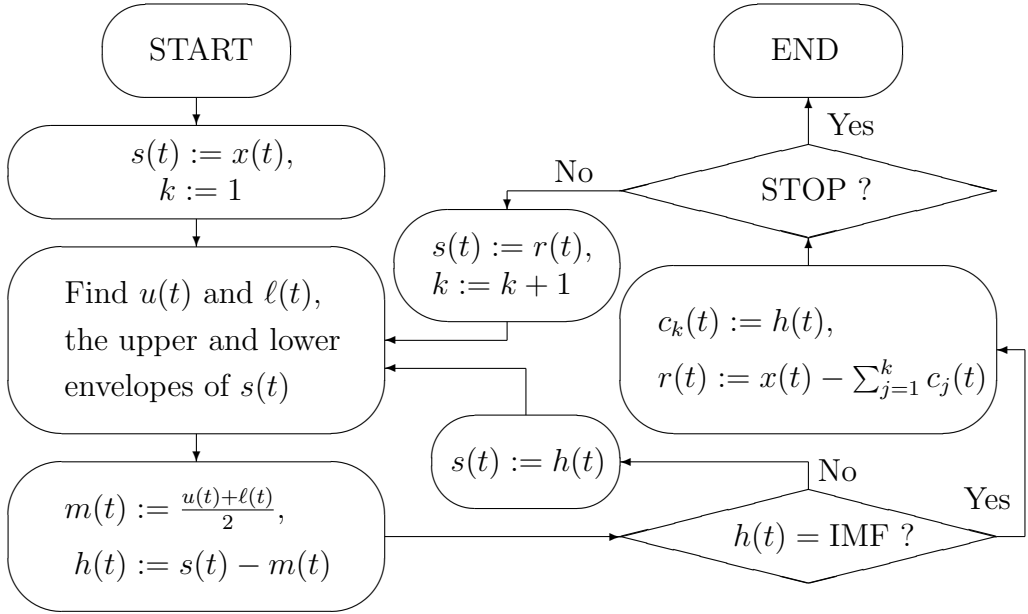

\section{Hilbert-Huang Transform and weighted sliding EMD}
\label{sec:hht}

The \ac{hht} is an extension of the Hilbert transform by adding the \ac{emd} to the Hilbert spectral analysis.
The Hilbert transform $\mathcal{H}[\cdot]$ is a linear operator used to obtain the analytic representation $\tilde{x}(t) = x(t) + i \mathcal{H}[x](t)$ of a real function $x(t)$,
\begin{equation}
  \label{eq:ht}
  \mathcal{H}[x](t) = \frac{1}{\pi} \mathcal{P} \int_{-\infty}^{\infty} \rmd \tau \frac{x(\tau)}{t - \tau},
\end{equation}
where $\mathcal{P}$ represents the principal value.
The Hilbert spectral analysis then finds the instantaneous amplitude $a(t) = || \tilde{x}(t) ||$ and the instantaneous frequency $f(t) = (2\pi)^{-1} (\rmd / \rmd t) \arg[\tilde{x}(t)]$ from the analytic representation.
In general, it is easy that the instantaneous amplitude and instantaneous frequency using the Hilbert spectral analysis fail to represent the accurate amplitude and frequency of the original time series $x(t)$.
For example, if the given function is shifted by a constant, $x(t) \to x(t) + x_0$, the analytic representation is also shifted by the same amount, $\tilde{x}(t) \to \tilde{x}(t) + x_0$, leading the contaminated instantaneous amplitude and instantaneous frequency. To resolve this contamination, the \ac{emd} is introduced to decompose the given function into several \acp{imf} that are well-behaved for the Hilbert spectral analysis.

\begin{figure}[tbp]
\begin{center}
\includegraphics[width=.8\textwidth]{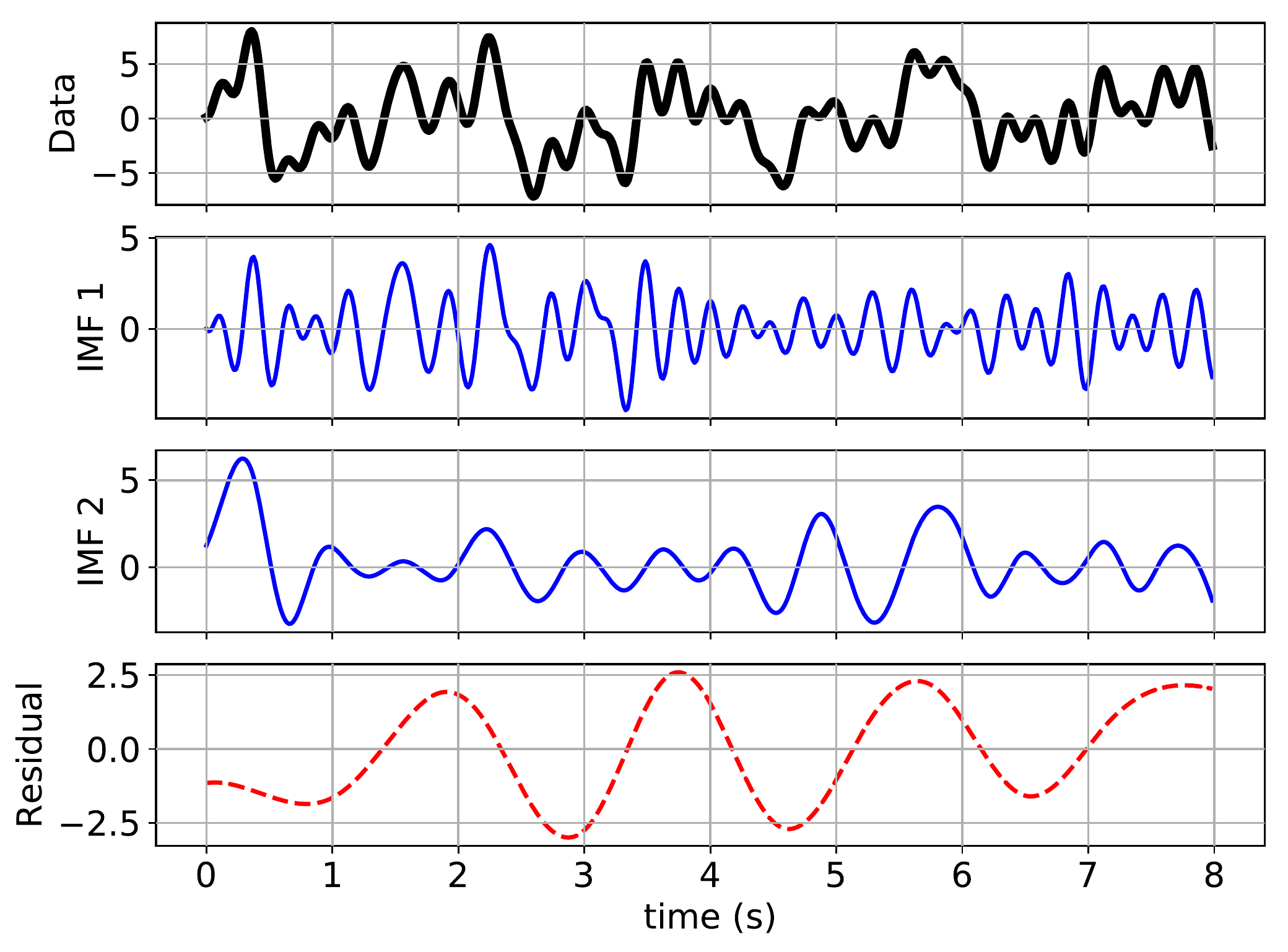}
\caption{An example of \ac{emd}. The data (top, black, thick) is decomposed into the \acsp{imf} (middle, bue) and the residual (bottom, red, dashed). The first \ac{imf} is dominated by high frequency fluctuation, while the lower frequency features reside in the second \ac{imf} and the residual.}
\label{fig:imf}
\end{center}
\end{figure}

The \ac{emd} is an adaptive process to obtain the \acp{imf} from a time-series data $x(t)$.
The algorithm of the \ac{emd} can be summarized as follows (see figure~\ref{fig:emd_flow}):
\begin{enumerate}
\item Set data to be sifted $s(t) := x(t)$ and the \ac{imf} number $k := 1$.
\item \label{emd:sifting} Find the upper and lower envelopes of $s(t)$, say $u(t)$ and $\ell(t)$.
\item \label{emd:subtraction} Subtract the mean of the envelopes $m(t) := \frac12 [u(t) + \ell(t)]$ from $s(t)$ and set $h(t) := s(t) - m(t)$.
\item \label{emd:imf} If $h(t)$ fails to satisfy a criterion for the \ac{imf} to be defined below, set $s(t) := h(t)$ and go to step~\ref{emd:sifting}.
\item \label{emd:nextimf} Otherwise, set $k$-th \ac{imf} $c_k(t) := h(t)$ and the residual $r(t) := x(t) - \sum_{j=1}^k c_j (t)$.
\item \label{emd} If $r(t)$ fails to satisfy ending criterion for the \ac{emd}, set $s(t) := r(t)$ and $k := k + 1$ then go to step~\ref{emd:sifting}.
\item Otherwise, finish the \ac{emd} with $k$ \acp{imf} $c_j(t)$ and the residual $r(t)$.
\end{enumerate}
In the literatures, the Cauchy type criterion~\cite{Huang1998}, the mean value criterion~\cite{Flandrin2004}, and the $S$-number criterion~\cite{Huang2003} are widely used for the step~\ref{emd:imf} (for a review, see~\cite{Wang2010}).
In the examples in this paper, the $S$-number criterion will be implicitly considered.
The $S$-number criterion restricts the difference between the numbers of extrema and zero-crossings of an \ac{imf}.
For the step~\ref{emd}, the criterion is given by the number of the \acp{imf} or the number of extrema of the residual $r(t)$.
An example of the \ac{emd} is shown in figure~\ref{fig:imf}.

\begin{figure}[tbp]
\begin{center}
\setlength{\unitlength}{3em}
\begin{picture}(8,6)
  \multiput(.3,5.8)(0,-.7){2}{\line(1,0){7.4}}
  \multiput(.5,5.8)(.5,0){15}{\line(0,-1){.1}}
  \multiput(.5,5.6)(.5,0){6}{\line(0,-1){.1}}
  \multiput(7.5,5.6)(-.5,0){6}{\line(0,-1){.1}}
  \put(3.3,5.35){input data}
  \multiput(.5,5.4)(.5,0){6}{\line(0,-1){.1}}
  \multiput(7.5,5.4)(-.5,0){6}{\line(0,-1){.1}}
  \multiput(.5,5.2)(.5,0){15}{\line(0,-1){.1}}
  \put(2.12,5.35){\textcolor{red}{$\odot$}}

  \multiput(.3,4.9)(0,-.7){5}{\line(1,0){7.4}}
  \multiput(.5,4.9)(2,0){4}{\line(0,-1){.7}}
  \multiput(1,4.2)(2,0){4}{\line(0,-1){.7}}
  \multiput(1.5,3.5)(2,0){4}{\line(0,-1){.7}}
  \multiput(2,2.8)(2,0){3}{\line(0,-1){.7}}
  \put(.8,4.45){segment 1}
  \put(1.3,3.75){segment 2}
  \put(1.8,3.05){segment 3}
  \put(2.3,2.35){segment 4}
  \put(2.8,4.45){segment 5}
  \put(3.3,3.75){segment 6}
  \put(3.8,3.05){\textcolor{gray}{segment 7}}
  \put(4.3,2.35){\textcolor{gray}{segment 8}}
  \multiput(2,4.9)(0,-.7){3}{\line(0,-1){.1}}
  \multiput(2,4.3)(0,-.7){3}{\line(0,-1){.1}}
  \multiput(2.5,4.2)(0,-.7){3}{\line(0,-1){.1}}
  \multiput(2.5,3.6)(0,-.7){3}{\line(0,-1){.1}}
  \multiput(3,4.9)(.5,0){3}{\line(0,-1){.1}}
  \multiput(3,4.3)(.5,0){3}{\line(0,-1){.1}}
  \multiput(3.5,4.2)(.5,0){2}{\line(0,-1){.1}}
  \multiput(3.5,3.6)(.5,0){2}{\line(0,-1){.1}}
  \multiput(3,3.5)(1,0){2}{\line(0,-1){.1}}
  \multiput(3,2.9)(1,0){2}{\line(0,-1){.1}}
  \multiput(3,2.8)(.5,0){2}{\line(0,-1){.1}}
  \multiput(3,2.2)(.5,0){2}{\line(0,-1){.1}}
  
  \textcolor{cyan}{
  \qbezier(3.5,2.8)(3.7,2.8)(4,3.15)
  \qbezier(4.5,3.5)(4.3,3.5)(4,3.15)
  \qbezier(4.5,3.5)(4.7,3.5)(5,3.15)
  \qbezier(5.5,2.8)(5.3,2.8)(5,3.15)
  \qbezier(4,2.1)(4.2,2.1)(4.5,2.45)
  \qbezier(5,2.8)(4.8,2.8)(4.5,2.45)
  \qbezier(5,2.8)(5.2,2.8)(5.5,2.45)
  \qbezier(6,2.1)(5.8,2.1)(5.5,2.45)
  }

  \put(2.5,1.45){average}
  \put(2.25,1.9){\textcolor{red}{\vector(0,-1){.9}}}
  \multiput(2.75,1.9)(.5,0){2}{\line(0,-1){.2}}
  \multiput(2.75,1.3)(.5,0){2}{\vector(0,-1){.3}}
  \put(3.75,1.9){\vector(0,-1){.9}}

  \multiput(4,2)(2,0){2}{\line(0,-1){.4}}
  \put(4.9,1.7){L}
  \put(4.7,1.8){\vector(-1,0){.7}}
  \put(5.3,1.8){\vector(1,0){.7}}
  \multiput(5,.9)(.5,0){2}{\line(0,1){.4}}
  \put(5.15,1){b}
  \put(4.8,1.1){\vector(1,0){.2}}
  \put(5.7,1.1){\vector(-1,0){.2}}

  \multiput(.3,.8)(0,-.7){2}{\line(1,0){7.4}}
  \multiput(.5,.8)(.5,0){15}{\line(0,-1){.1}}
  \multiput(.5,.6)(.5,0){5}{\line(0,-1){.1}}
  \multiput(7.5,.6)(-.5,0){5}{\line(0,-1){.1}}
  \put(2.95,.35){averaged \acsp{imf}}
  \multiput(.5,.4)(.5,0){5}{\line(0,-1){.1}}
  \multiput(7.5,.4)(-.5,0){5}{\line(0,-1){.1}}
  \multiput(.5,.2)(.5,0){15}{\line(0,-1){.1}}
  \put(2.12,.35){\textcolor{red}{$\odot$}}
\end{picture}
\caption{An example of weighted \acs{semd} for $N=4$. The length of each segment is $L=Nb$, where $b$ is the length of a block. The segments $n$ and $(n+1)$ are overlapped by the length of $(N-1)b$ and each block is shared by four segments. For example, the block marked with \textcolor{red}{$\odot$} is shared by segments 1, 2, 3 and 4. The \acs{emd} process decomposes each segment into the corresponding \acsp{imf}, which will then be multiplied by a weighting function. The weighting function is shown in segments 7 and 8. Then, the \acsp{imf} of all the segments in a shared block are averaged out with the weighting function. For the block \textcolor{red}{$\odot$}, the \acsp{imf} in the last block of the segment 1, in the second last of 2, in the second of 3 and in the first of 4 are weighted-averaged. When the weighting function is a nonzero constant function, the weighted \acs{semd} reduces to the \acs{semd}.}
\label{fig:wsemd}
\end{center}
\end{figure}
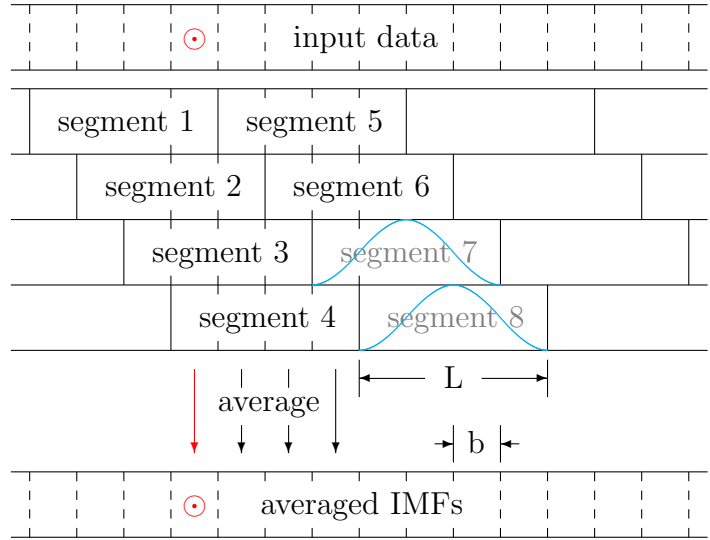

The \ac{emd} approach can be extended in a number of ways, some of which are important here so that they help to improve the accuracy of decomposition and to suppress the computation cost. The ensemble \ac{emd} is introduced in order to obtain more accurate \acp{imf}~\cite{Wu2009}. It applies step~\ref{emd:sifting} of \ac{emd} to an ensemble of whitened data and their averaged value as a true result, which reduces the noise perturbation around the true solution and enhance the precision of \acp{imf}.
In general, however, it is inappropriate to use the ensemble \ac{emd} in an online data analysis because of its expensive computing cost.
This drawback can be also resolved by the \ac{semd}~\cite{Faltermeier2010} and/or its extension, weighted \ac{semd}~\cite{Faltermeier2011}.
The algorithm of \ac{semd} is somewhat simple (see figure~\ref{fig:wsemd}):
\begin{enumerate}
\item The input data is split into segments of length $L$, where each segment can be split into $N$ blocks of length $b=L/N$ and is overlapped with the next segment by $(N-1)$ blocks, that is, each block is shared by $N$ segments.
\item Each segment is decomposed by the \ac{emd} process.
\item \label{semd:avg} The \acp{imf} of the segments are averaged out in each block.
\end{enumerate}
The weighted \ac{semd} introduces a \emph{weighting function} on averaging process in step~\ref{semd:avg}, which reduces the edge effects at both sides of each segment originated from the \ac{emd} and from the \ac{semd} as well.
The authors in~\cite{Faltermeier2011} suggest a Gaussian weighting function, $w(n) = \exp \left\{ \frac12 [2 \alpha n / (L-1)]^2 \right\}$ with $\alpha=2.5$ and $-\frac12 (L-1) \le n \le \frac12 (L-1)$.
This function, however, does not vanish at the ends of a segment, i.e. $n = \pm \frac12 (L-1)$, and thus the edge effect from the \ac{semd} remains and causes a step function-like behavior in the \acp{imf}.
To moderate this unwanted behavior, we, here, propose a sinusoidal weighting function,
\begin{equation}
  w(n) = \sin^2 \left( \frac{\pi n}{L-1} \right), \qquad n = 0, \cdots, (L-1),
\end{equation}
which vanishes at the edges and thus produces well-behaved \acp{imf} in each segment.

Once the data $x(t)$ is decomposed into \acp{imf} $c_j(t)$, the analytic representation of each \ac{imf} is calculated by the Hilbert transform~\eqref{eq:ht} to obtain its instantaneous amplitude and instantaneous frequency, say $a_j(t)$ and $f_j(t)$.
Then, we can recover the original data $x(t)$ in terms of the instantaneous amplitudes and the instantaneous frequencies of the \acp{imf},
\begin{equation}
  \label{eq:hht}
  x(t) = \Re \left\{ \sum_j a_j(t) \exp \left[ 2\pi \rmi \int^t f_j(\tau) \rmd \tau \right] \right\} + r(t),
\end{equation}
where $\Re(z)$ represents the real part of $z$. The residual $r(t)$ is negligible in the analysis, since it contains in general only the lower frequency data than the frequency band under consideration.
Note that the representation~\eqref{eq:hht} of the \ac{hht} is similar to the discrete Fourier transform $x(t) = \Re \left[ \sum_j a_j \exp(2\pi \rmi f_j t) \right]$ but the $a_j(t)$ and the $f_j(t)$ are not constant, which explains why the \ac{hht} has much smaller number of modes than the discrete Fourier transform because of the adaptive nature of the transform.

\section{Event trigger generation algorithm of \etagen{}}
\label{sec:etg}

\subsection{EtaGen Algorithm I: Finding Excesses}
We assume that the given data $x(t)=n(t)+g(t)$ is whitened, in order to make the event trigger generation algorithm more or less simple. Here we define $x(t)$ as the data, $n(t)$ as the background noise, and $g(t)$ as transients.
Then, let $\sigma$ be the \ac{std} of the background noise, which definitely follows the Gaussian distribution.
Since the data $x(t)$ contains both the background noise and the transients, it is hard to estimate $\sigma$.
We thus assume that the \ac{std} of the background Gaussian noise is proportional to the \ac{mad} of $x(t)$,
\begin{equation}
\label{MAD}
  \sigma = \gamma \, \textrm{MAD} \left[ x(t) \right],
\end{equation}
where the coefficient $\gamma$ is chosen to be $\madratio$ due to the fact that the ratio between the \ac{std} and the \ac{mad} of a Gaussian noise is given by $\left[ \sqrt{2}\, \textrm{erf}^{-1} (\frac12) \right]^{-1} \approx 1.48$.
Note that $x(t)$ contains transients in general and its \ac{mad} is thus larger than the \ac{mad} of the background Gaussian noise; furthermore, the \ac{std} and the \ac{mad} of a sine wave data are the same.
However, if the Gaussian noise is dominant in the data $x(t)$, then the contributions of the transients to the \ac{mad} are negligible.

\begin{figure}[tbp]
\begin{center}
\includegraphics[width=0.8\textwidth]{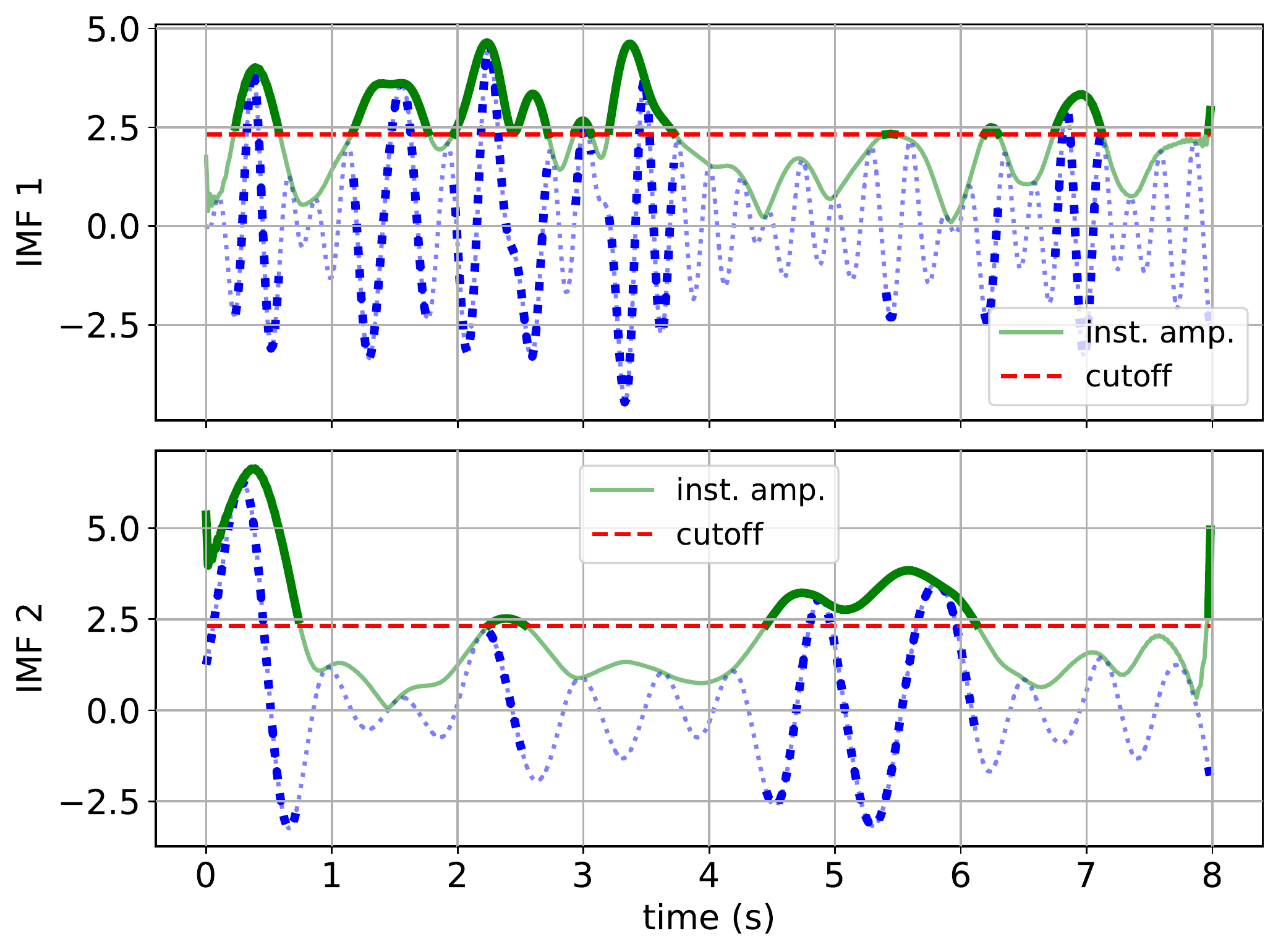}
\caption{This plots show how to find \utrgs{} in the \etagen{}.  First, series of maxima in the instantaneous amplitude (thick green solid line) are identified in every \ac{imf} by putting cutoffs (red dashed line) in amplitude. The \utrgs{} are the segments of time series in \acp{imf} corresponding to the series of maxima identified in the \acp{imf} (thick blue dotted line). The information of \autrg{} such as the peak amplitude, the peak time, the centreal frequency, the starting and end times and \acs{snr} is recorded if the \acs{snr} of the \utrg{} computed by~(\ref{eq:snr}) exceeds a given threshold, $T$. Refer to the text for more details. }
\label{fig:etg}
\end{center}
\end{figure}

To find \utrgs{}, we introduce $A_j$, a set of maxima of the instantaneous amplitude $a_j(t)$ of the $j$-th \ac{imf} $c_j(t)$.
Then, setting \iac{snr} threshold $T$, the \utrg{} finding algorithm for the $j$-th \ac{imf} is as follows:
\begin{enumerate}
\item Set a cutoff $\sigma_j := \min\{ \sigma, \textrm{MAD}[c_j(t)] \}$ and $k:=1$.
\item \label{trg:peak_amp} Take $k$-th maximum $A_{j,k}$ of the $j$-th instantaneous amplitude $a_j(t)$.
\item If $A_{j,k} < \sigma_j$, then set $k := k+1$ and go to step~\ref{trg:peak_amp}. An example of maxima over the cutoff is depicted in figure~\ref{fig:etg}.
\item \label{trg:tste} Otherwise, find the start and end times of the \utrg{}, say $t_\mathrm{s}$ and $t_\mathrm{e}$, where we omit the indices $j$ and $k$ for convenience.
\item Calculate the \ac{snr} $\rho$ of the \utrg{}.
\item If $\rho \ge T$, then store the \utrg{} informations, such as the peak amplitude $A_{j,k}$, the peak time $t_\mathrm{p}$ where $A_{j,k}$ is located, the peak frequency $f_j(t_\mathrm{p})$, the central time $t_\mathrm{c} = \mathcal{N} \int_{t_\mathrm{s}}^{t_\mathrm{e}} t \, a_j(t) \, \rmd t$, the central frequency $f_\mathrm{c} = \mathcal{N} \int_{t_\mathrm{s}}^{t_\mathrm{e}} f_j(t) \, a_j(t) \, \rmd t$, etc.\ as well as $t_\mathrm{s}$, $t_\mathrm{e}$, and $\rho$, where $\mathcal{N} = \left[ \int_{t_\mathrm{s}}^{t_\mathrm{e}} a_j(t) \, \rmd t \right]^{-1}$.
\item If $k$ is less than the number of the maxima $A_j$, then set $k := k+1$ and go to step~\ref{trg:peak_amp}.
\item Otherwise, finish finding \utrgs{} for the $j$-th \ac{imf} and move on to the next \ac{imf} if exists.
\end{enumerate}
To find $t_\mathrm{s}$ and $t_\mathrm{e}$ in the step~\ref{trg:tste}, one might consider the use of the instantaneous amplitude: the times, when the nearest minima of $a_j(t)$ to the peak amplitude $A_{j,k}$ are located, can be set to $t_\mathrm{s}$ and $t_\mathrm{e}$.
However, the instantaneous amplitude is sometimes obtained as a sawtooth shape instead of the smooth bell shape for \autrg{}.
To resolve it, we introduce the extremum $c^{(\mathrm{ext})}_j$ of the $j$-th \ac{imf} $c_j(t)$, find the nearest minima of $|c^{(\mathrm{ext})}_j|$ to the peak amplitude $A_{j,k}$, and set $t_\mathrm{s}$ and $t_\mathrm{e}$ times when they are located.
We then calculate the \ac{snr} of the \utrg{} corresponding to the peak amplitude $A_{j,k}$ in the $j$-th \ac{imf} as
\begin{equation}
  \label{eq:snr}
  \rho_{j,k} = \frac{1}{\sigma} \left[ \int_{t_\mathrm{s}}^{t_\mathrm{e}} | c_j(t) |^2 \, \rmd t \right]^{1/2},
\end{equation}
which agrees with the \ac{snr} definition shown in~\cite{Stroeer:2009zz}.

\begin{figure}[tbp]
\begin{center}
\includegraphics[width=0.8\textwidth]{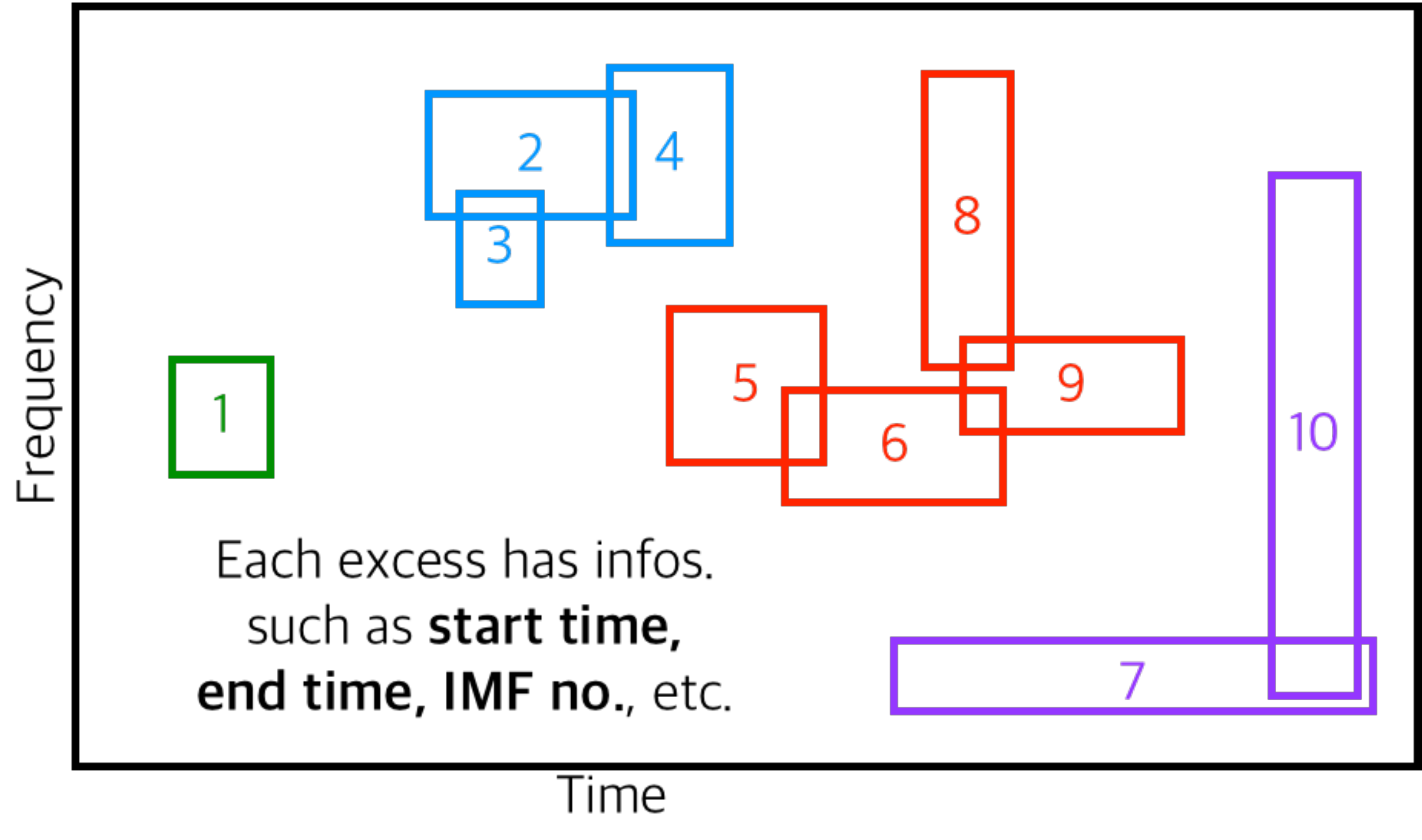}
\caption{Illustration of clustering \utrgs{}. Each rectangle represents \autrg{} in the time-frequency plane. It is defined by the starting and end time ($t_\mathrm{s}$ and $t_\mathrm{e}$, respectively) and the low and high frequency ($f_\mathrm{low}$ and $f_\mathrm{high}$, respectively) of the \utrg{}. Two or more rectangles are clustered in to a single \trg{} if they overlap.
For example, red rectangles are clustered into a single \trg{}. And so are the light blue and pupple ones.}
\label{fig:cluster}
\end{center}
\end{figure}

\subsection{EtaGen Algorithm II: Clustering}

A transient event signal as well as background noises should be split into several modes of \ac{imf}, which generates a set of \utrgs{}.
We now gather such \utrgs{} together to get \emph{\atrg{}}.
Among the parameters of each \utrg{}, the start time $t_\mathrm{s}$, the end time $t_\mathrm{e}$, the high frequency $f_\mathrm{high}$ and the low frequency $f_\mathrm{low}$ determine a square section in time-frequency plane, as seen in figure~\ref{fig:cluster}.
The high and low frequencies are selected as the upper and lower quartiles in the frequency distribution of the corresponding \ac{imf} between $t_\mathrm{s}$ and $t_\mathrm{e}$.
Each square in figure~\ref{fig:cluster} represents \autrg{} and the overlapped \utrgs{} are clustered into \atrg{}.
The squares may be stretched to reflect some tolerances in time and/or frequency.
If another \utrg{} is overlapped one of the \utrgs{} in \atrg{}, then it should be clustered into the same \trg{}.
For example, in figure~\ref{fig:cluster}, the \utrgs{} 5 and 6 are overlapped each other and so they form \atrg{}.
The \utrgs{} 7 and 8 are overlapped with neither of 5 and 6, so they are not clustered in the same \trg{} at the moment.
After the \utrg{} 9 is clustered in the \trg{} since it is overlapped with 6, the \utrg{} 8 is now overlapped with 9 and is clustered into the \trg{}.
The \utrg{} 7 overlaps with 10 to form the last \trgs{} in this example.
There are four \trgs{} in figure~\ref{fig:cluster}: one with the \utrg{} 1, another with 2, 3 and 4, and the others with 5, 6, 8 and 9 and with 7 and 10.
The peak time, peak frequency, and peak amplitude for \atrg{} are chosen as those for the \utrg{} that has the largest peak amplitude among the \utrgs{} in the cluster.

\begin{figure}[tbp]
\begin{center}
\includegraphics[width=0.8\textwidth]{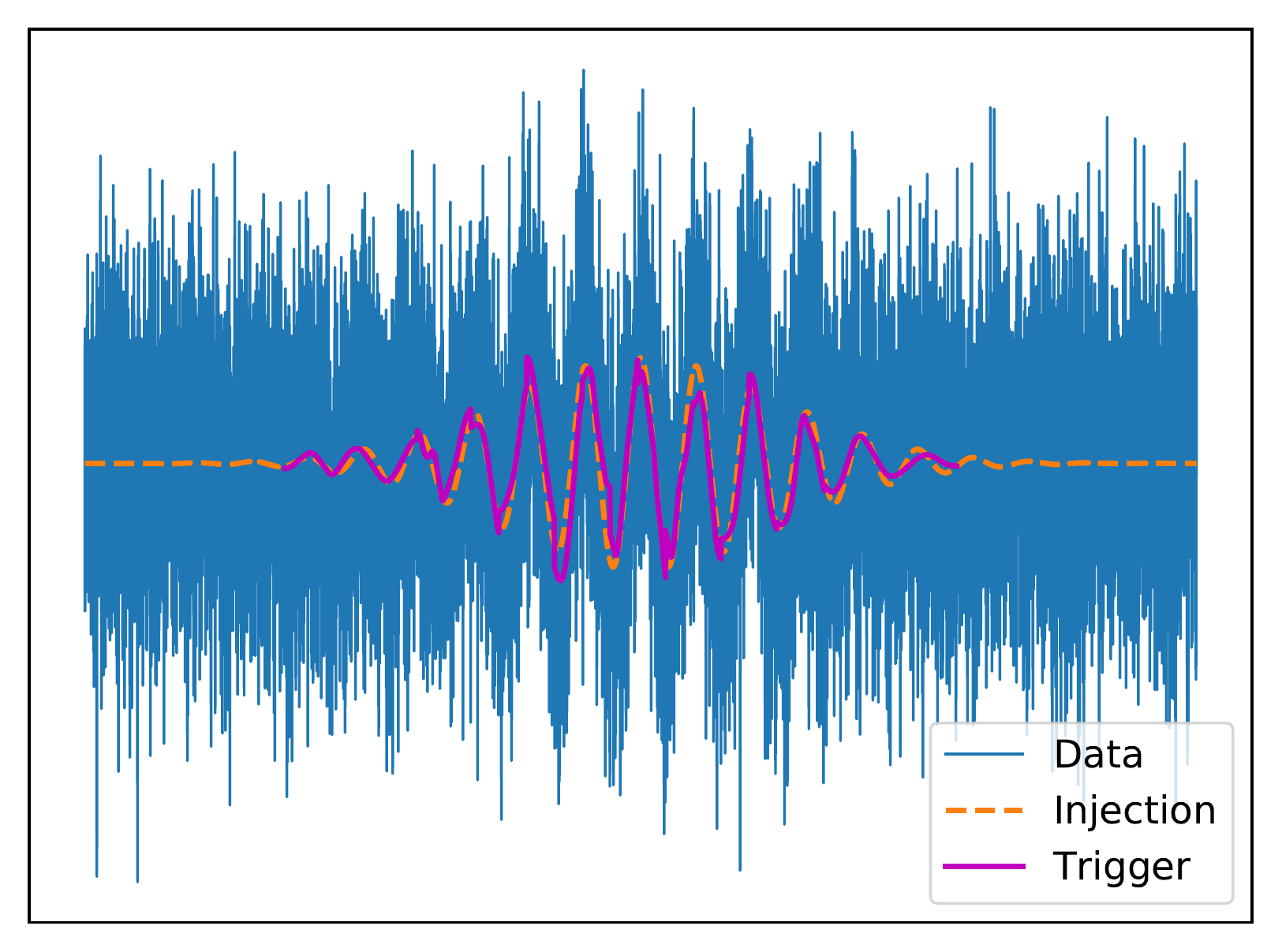}
\caption{Waveform reconstruction of a sine-Gaussian injection (orange, dashed). \Atrg{}'s waveform (purple) can be reconstructed by superposing all \utrgs{} that comprise the \trg{}. We compute \ac{snr} of the \trg{} using its reconstructed waveform by applying~(\ref{eq:snr_trg}). The thin blue line represents the original data that contains sine-Gaussian waveform with noise. Note that the reconstructed waveform  recovers the injected sine-Gaussian waveform with small discrepancy.}
\label{fig:waveform}
\end{center}
\end{figure}

\subsection{Signal-to-Noise Ratio from Waveform Reconstruction}

\Atrg{} consists of one or more \utrgs{} and each \utrg{} can be restored in the time domain, i.e. the corresponding \ac{imf} between $t_\mathrm{s}$ and $t_\mathrm{e}$. Therefore, by superposing the time-series of the \utrgs{} in the cluster, the original waveform of the event can be reconstructed. One can calculate the \ac{snr} of the \trg{} using this reconstructed waveform as
\begin{equation}
  \rho = \frac{1}{\sigma} \left[ \int_{T_\mathrm{s}}^{T_\mathrm{e}} | X(t) |^2 \, \rmd t \right]^{1/2},
\label{eq:snr_trg}
\end{equation}
where $T_\mathrm{s} = \min \{t_\mathrm{s}\}$ and $T_\mathrm{e} = \max \{t_\mathrm{e}\}$ are the start and end times of the \trg{}, respectively, and $X(t)$ is the reconstructed waveform by superposing the \utrgs{}.

The \ac{snr} computed from the reconstructed waveform is \emph{little} higher than that computed from~(\ref{eq:snr}). However, the accuracy of the \ac{snr} recovery depends on how precisely superposed waveform is reconstructed, i.e. how accurate the decomposed \acp{imf} are. To enhance the accuracy of the recovery, one can performs more precise algorithm of finding \utrgs{} and clustering but it always produces a trade-off with computation time. Indeed, the overcome of uncertainty between time-frequency resolution might be transferred to the uncertainty between definite algorithm and computation time.

\section{The performance of the \etagen{}: Comparison to Omicron}
\label{sec:perf}
In this section, the performance of the \etagen{} is compared to that of the Omicron based on \emph{Q-transform} which is \emph{currently} the best performing event trigger generator used in the \ac{gw} data analysis~\cite{McIver:2015pms}. 

The performance of the \etagen{} may be investigated quantitatively by observing how many artificial transients injected in a background noise can be found. 
We generated simulated Gaussian noise colored to have the same noise curve as the \ac{aligo} detectors at design sensitivity. We then injected sine-Gaussians and white noise bursts 30 seconds apart as target signals for recovery.
The total number of sine-Gaussian (white noise burst) injections is \sgtotal{} (\wnbtotal{}). The injections are almost uniformly distributed in both \ac{snr} and frequency.

We set the \ac{snr} threshold for \trgs{} (\utrgs{}) in the \etagen{} to \snrth{} (\usnrth{}). One can increase the number of \utrgs{} by lowering the \ac{snr} threshold. However, if we have a excessively large number of \utrgs{} by decreasing the threshold to a very small value, the clustering algorithm produces event triggers with unrealistically long duration which are obviously far from the real event triggers. We empirically choose the \ac{snr} threshold value of \usnrth{} for \utrg{} to achieve balance between the number of \utrgs{} and the duration of \emph{the clustered} event triggers after running many tests.

\begin{figure}[tbp]
\begin{center}
\includegraphics[width=0.8\textwidth]{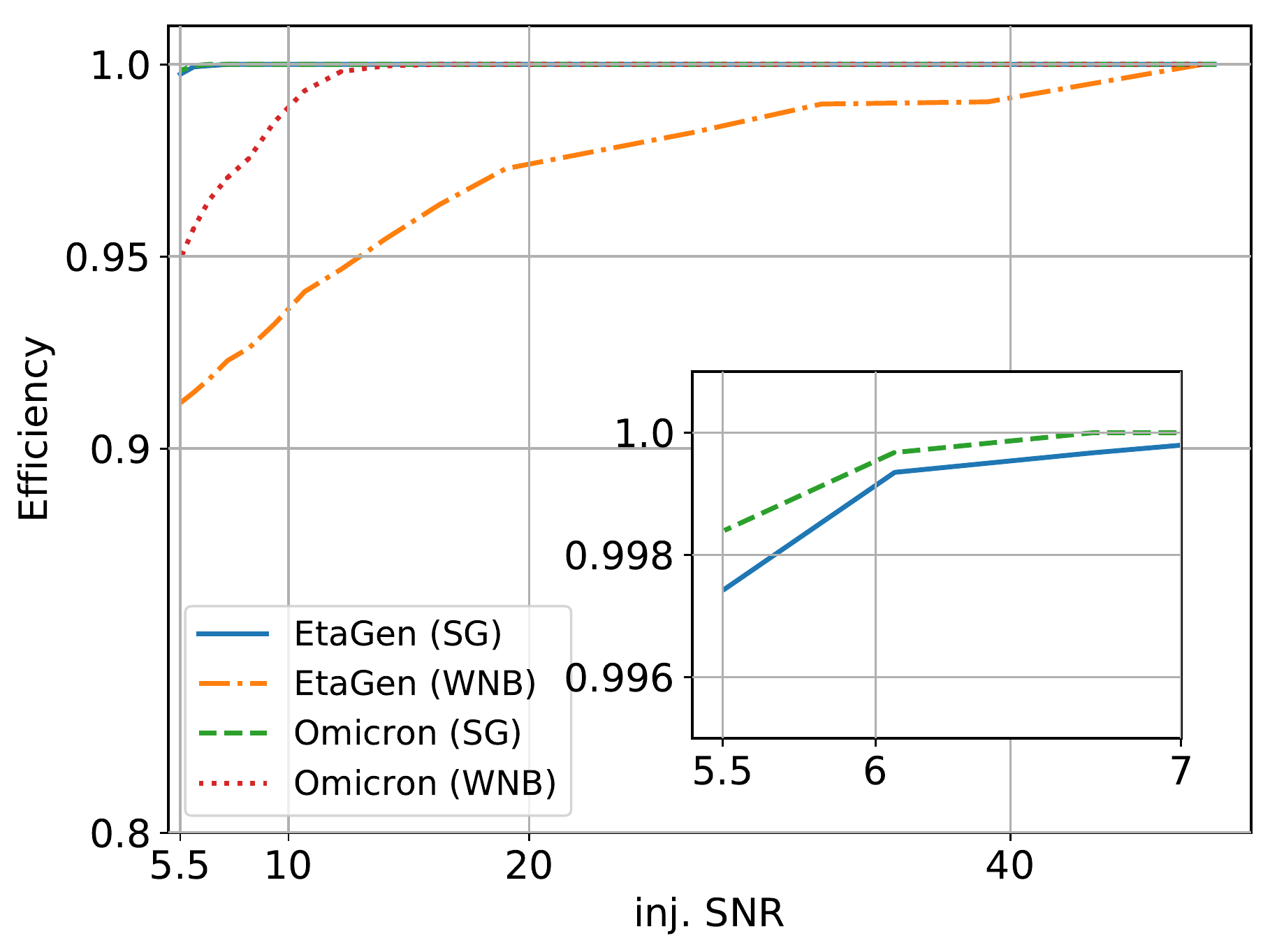}
\caption{The comparison of event trigger generator performance between \etagen{} and Omicron for sine-Gaussian (SG) and white noise burst (WNB) injections}
\label{fig:roc}
\end{center}
\end{figure}

Consequently, \etagen{} finds in total \allpercent{} of the injections; \sgpercent{} for sine-Gaussian and \wnbpercent{} for white noise burst, respectively. As for the timing accuracy, the \ac{std} of the timing error of all injections is \alltacc{}; sine-Gaussian \sgtacc{} and white noise burst \wnbtacc{}.
The \ac{std} of the frequency error of all injections is \allfacc{}; sine-Gaussian \sgfacc{} and white noise burst \wnbfacc{}.

One can compare these results to the performance from Omicron, having efficiency of $99.80\,\%$ (sine-Gaussian) and $95.20\,\%$ (white noise burst), timing accuracy of $3.4\times10^{-4}\,\textrm{s}$ (sine-Gaussian) and $9.8\times 10^{-3}\,\textrm{s}$ (white noise burst), and frequency accuracy of $82\,\textrm{Hz}$ (sine-Gaussian) and $152\,\textrm{Hz}$ (white noise burst).
We conclude that the efficiency for finding event triggers of \etagen{} is comparable with the Omicron. However, the timing and frequency accuracies of triggers are worse than those of Omicron for sine-Gaussian (white noise burst) waveform. 
The results of the event trigger generator performance tests and comparison to Omicron are seen in figure~\ref{fig:roc}, and detailed results are summarized in table~\ref{tab:snr}.

\begin{table*}[tbp]
\caption{The event trigger generator performance for \etagen{} and Omicron with \acs{snr} threshold 5.5 is shown.
The numbers in the efficiency and timing/frequency accuracies are given by [(number of found injections)/(total number of injections)]$\times$100\,\%, std[(trigger time)$-$(injection time)], and std[(trigger frequency)$-$(injection frequency)], respectively.
}
\label{tab:snr}
\begin{indented}
\lineup
\item[]\begin{tabular}{@{}*{4}{l}}
  \br
  \multicolumn{2}{c}{\multirow{2}{*}{Performance tests}} & {\etagen{}} & {Omicron}
  \\
  \multicolumn{2}{c}{} & SNR threshold 5.5  & SNR threshold 5.5
  \\
  \mr
  \multirow{2}{*}{Efficiency} & sine-Gaussian & \0\sgpercent{} & \099.8\,\%
  \\
  & white noise burst & \0\wnbpercent{} & \095.2\,\%
  \\
  \multirow{2}{*}{Timing accuracy} & sine-Gaussian & \0\0\sgtacc{} & \0\03.4$\times 10^{-4}$\,s
  \\
  & white noise burst & \0\0\wnbtacc{} & \0\09.8$\times 10^{-3}$\,s
  \\
  \multirow{2}{*}{Frequency accuracy} & sine-Gaussian & \sgfacc{}  & \082\,Hz
  \\
  & white noise burst & \wnbfacc{} & 152\,Hz
  \\
  \br
\end{tabular}
\end{indented}
\end{table*}

\section{Discussion}
\label{sec:dis}

We have developed \etagen{}, a new event trigger generator based on \ac{hht} utilizing the adaptive nature of the transform. In order to evaluate its performance, we compare the efficiency and accuracy of \etagen{} with those of Omicron by running the event trigger generator on the \ac{aligo} simulated data with two different types of injected signals; sine-Gaussian and white noise burst. As summarized in table~\ref{tab:snr}, the result shows that the efficiency of \etagen{} in finding event triggers are comparable to that of Omicron, which is the first step of demonstration as a viable event trigger generator.

The inaccuracy in timing and frequency can arise from errors in weighted \ac{semd} due to parameters chosen such as $N$, $L$ and the weighting function. Increasing $N$ and $L$ might help to obtain more precise \acp{imf} because it corresponds to increasing the ensemble size, although the decomposition process becomes computationally expensive. More sophisticated \utrg{} finding procedure using better \ac{snr} calculation and an alternative cutoff criterion (see, e.g., \cite{rousseeuw-qn-1993}) can also improve the accuracy. In fact, we observe improvement in the accuracies and efficiency of recovering triggers by changing \ac{snr} calculation from~(\ref{eq:snr}) to~(\ref{eq:snr_trg}) increases.

As stated at the beginning of this paper, it is anticipated that \etagen{} will show better performance in recovering low frequency triggers because of the adaptive nature of the mode decomposition. The time-frequency accuracy trade-off that appears in other event trigger generators utiliziing fixed harmonic bases for decomposition such as Fourier transform and wavelet transform is expected to be less severe in \etagen{}. This is under investigation and will be reported in a separate article.

\ack
The authors would like to thank Jay Tasson for helpful comments and suggestions and Chris Pankow for letting us to use the simulated GW data.
EJS is grateful to Hyoungseok Chu for the fruitful contribution to the early stage of this work.
YMK was supported by the National Research Foundation of Korea (NRF) grant funded by the 
Korea government (MSIP) (No.\ 2016R1A5A1013277).
LIGO was constructed by the California Institute of Technology and Massachusetts Institute of Technology with funding from the National Science Foundation and operates under cooperative agreement PHY-0757058 . This paper carries LIGO Document Number LIGO-P1800294.

\section*{References}
\bibliographystyle{unsrt}
\bibliography{Etagen}

\begin{thebibliography}{10}

\bibitem{TheLIGOScientific:2016wyq}
B.~P. Abbott et~al.
\newblock {GW150914: Implications for the stochastic gravitational wave
  background from binary black holes}.
\newblock {\em Phys. Rev. Lett.}, 116(13):131102, 2016.

\bibitem{Abbott:2016nmj}
B.~P. Abbott et~al.
\newblock {GW151226: Observation of Gravitational Waves from a 22-Solar-Mass
  Binary Black Hole Coalescence}.
\newblock {\em Phys. Rev. Lett.}, 116(24):241103, 2016.

\bibitem{Abbott:2017vtc}
Benjamin~P. Abbott et~al.
\newblock {GW170104: Observation of a 50-Solar-Mass Binary Black Hole
  Coalescence at Redshift 0.2}.
\newblock {\em Phys. Rev. Lett.}, 118(22):221101, 2017.

\bibitem{Abbott:2017gyy}
B..~P.. Abbott et~al.
\newblock {GW170608: Observation of a 19-solar-mass Binary Black Hole
  Coalescence}.
\newblock {\em Astrophys. J.}, 851(2):L35, 2017.

\bibitem{Abbott:2017oio}
B.~P. Abbott et~al.
\newblock {GW170814: A Three-Detector Observation of Gravitational Waves from a
  Binary Black Hole Coalescence}.
\newblock {\em Phys. Rev. Lett.}, 119(14):141101, 2017.

\bibitem{TheLIGOScientific:2017qsa}
Benjamin~P. Abbott et~al.
\newblock {GW170817: Observation of Gravitational Waves from a Binary Neutron
  Star Inspiral}.
\newblock {\em Phys. Rev. Lett.}, 119(16):161101, 2017.

\bibitem{0264-9381-32-24-245005}
LÂK Nuttall, T~J Massinger, J~Areeda, J~Betzwieser, S~Dwyer, A~Effler, R~P
  Fisher, P~Fritschel, J~S Kissel, A~P Lundgren, D~M Macleod, D~Martynov,
  J~McIver, A~Mullavey, D~Sigg, J~R Smith, G~Vajente, A~R Williamson, and C~C
  Wipf.
\newblock Improving the data quality of advanced ligo based on early
  engineering run results.
\newblock {\em Classical and Quantum Gravity}, 32(24):245005, 2015.

\bibitem{TheLIGOScientific:2016zmo}
B.~P. Abbott et~al.
\newblock {Characterization of transient noise in Advanced LIGO relevant to
  gravitational wave signal GW150914}.
\newblock {\em Class. Quant. Grav.}, 33(13):134001, 2016.

\bibitem{Huang:2005:book}
Norden~E Huang and Samuel S~P Shen.
\newblock {\em Hilbert-Huang Transform and Its Applications}.
\newblock WORLD SCIENTIFIC, 2005.

\bibitem{Huang1998}
Norden~E. Huang, Zheng Shen, Steven~R. Long, Manli~C. Wu, Hsing~H. Shih, Quanan
  Zheng, Nai-Chyuan Yen, Chi~Chao Tung, and Henry~H. Liu.
\newblock The empirical mode decomposition and the hilbert spectrum for
  nonlinear and non-stationary time series analysis.
\newblock {\em Proc.\ R.\ Soc.\ Lond.\ A}, 454(1971):903--995, 1998.

\bibitem{Huang2008}
Norden~E. Huang and Zhaohua Wu.
\newblock A review on hilbert-huang transform: Method and its applications to
  geophysical studies.
\newblock {\em Reviews of Geophysics}, 46(2):RG2006, 2008.

\bibitem{Camp:2007ee}
Jordan~B. Camp, John~K. Cannizzo, and Kenji Numata.
\newblock {Application of the Hilbert-Huang Transform to the Search for
  Gravitational Waves}.
\newblock {\em Phys. Rev.}, D75:061101, 2007.

\bibitem{Stroeer:2009zz}
Alexander Stroeer, John~K. Cannizzo, Jordan~B. Camp, and Nicolas Gagarin.
\newblock {Methods for detection and characterization of signals in noisy data
  with the Hilbert-Huang transform}.
\newblock {\em Phys. Rev.}, D79:124022, 2009.

\bibitem{Stroeer:2009hv}
Alexander Stroeer and Jordan Camp.
\newblock {Ninja data analysis with a detection pipeline based on the
  Hilbert-Huang Transform}.
\newblock {\em Class. Quant. Grav.}, 26:114012, 2009.

\bibitem{Kaneyama:2016fww}
Masato Kaneyama, Ken-ichi Oohara, Hirotaka Takahashi, Yuichiro Sekiguchi,
  Hideyuki Tagoshi, and Masaru Shibata.
\newblock {Analysis of gravitational waves from binary neutron star merger by
  Hilbert-Huang transform}.
\newblock {\em Phys. Rev.}, D93(12):123010, 2016.

\bibitem{Sakai:2017ckm}
Kazuki Sakai, Ken-Ichi Oohara, Hiroyuki Nakano, Masato Kaneyama, and Hirotaka
  Takahashi.
\newblock {Estimation of starting times of quasi-normal modes in ringdown
  gravitational waves with the Hilbert-Huang transform}.
\newblock {\em Phys. Rev.}, D96(4):044047, 2017.

\bibitem{Valdes:2017xce}
Guillermo Valdes, Brian O'Reilly, and Mario Diaz.
\newblock {A Hilbert–Huang transform method for scattering identification in
  LIGO}.
\newblock {\em Class. Quant. Grav.}, 34(23):235009, 2017.

\bibitem{Flandrin2004}
P.~Flandrin, G.~Rilling, and P.~Goncalves.
\newblock Empirical mode decomposition as a filter bank.
\newblock {\em IEEE Signal Processing Letters}, 11(2):112--114, Feb 2004.

\bibitem{Huang2003}
Norden~E Huang, Man-Li~C Wu, Steven~R Long, Samuel~S.P Shen, Wendong Qu, Per
  Gloersen, and Kuang~L Fan.
\newblock A confidence limit for the empirical mode decomposition and hilbert
  spectral analysis.
\newblock {\em Proc.\ R.\ Soc.\ Lond.\ A}, 459(2037):2317--2345, 2003.

\bibitem{Wang2010}
Gang Wang, Xian-Yao Chen, Fang-Li Qiao, Zhaohua Wu, and Norden~E. Huang.
\newblock On intrinsic mode function.
\newblock {\em Adv. Adapt. Data Anal.}, 02(03):277--293, 2010.

\bibitem{Wu2009}
Zhaohua Wu and Norden~E. Huang.
\newblock Ensemble empirical mode decomposition: A noise-assisted data analysis
  method.
\newblock {\em Adv. Adapt. Data Anal.}, 01(01):1--41, 2009.

\bibitem{Faltermeier2010}
R.~Faltermeier, A.~Zeiler, I.~R. Keck, A.~M. Tomé, A.~Brawanski, and E.~W.
  Lang.
\newblock Sliding empirical mode decomposition.
\newblock In {\em The 2010 International Joint Conference on Neural Networks
  (IJCNN)}, pages 1--8, July 2010.

\bibitem{Faltermeier2011}
R.~Faltermeier, A.~Zeiler, A.~M. Tom\'e, A.~Brawanski, and E.~W. Lang.
\newblock Weighted sliding empirical mode decomposition.
\newblock {\em Adv. Adapt. Data Anal.}, 03(04):509--526, 2011.

\bibitem{McIver:2015pms}
Jessica~L. McIver.
\newblock {\em {The impact of terrestrial noise on the detectability and
  reconstruction of gravitational wave signals from core-collapse supernovae}}.
\newblock PhD thesis, Massachusetts U., Amherst, 2015.

\bibitem{rousseeuw-qn-1993}
Peter~J. Rousseeuw and Christophe Croux.
\newblock Alternatives to the median absolute deviation.
\newblock {\em Journal of the American Statistical Association}, 88(424), 1993.

\end{thebibliography}

\end{document}